\PassOptionsToPackage{unicode}{hyperref}
\PassOptionsToPackage{hyphens}{url}
\PassOptionsToPackage{dvipsnames,svgnames,x11names}{xcolor}
\documentclass[
  12pt]{article}

\usepackage[margin=1.3in]{geometry}
\usepackage{amsmath}
\usepackage{amsthm}
\usepackage{xr-hyper}
\usepackage{hyperref}
\usepackage{natbib}
\usepackage{bbold}
\usepackage{caption}
\usepackage{subcaption}
\usepackage{booktabs}
\usepackage{array}
\usepackage{placeins}
\usepackage{xcolor}
\usepackage{graphicx}
\usepackage{comment}
\usepackage{multirow}
\usepackage{enumitem}
\usepackage{pdfpages}
\usepackage[linesnumbered,ruled,vlined]{algorithm2e}

\newtheorem{prop}{Proposition}

\usepackage{amssymb}
\usepackage{iftex}
\ifPDFTeX
  \usepackage[T1]{fontenc}
  \usepackage[utf8]{inputenc}
  \usepackage{textcomp} 
\else 
  \usepackage{unicode-math}
  \defaultfontfeatures{Scale=MatchLowercase}
  \defaultfontfeatures[\rmfamily]{Ligatures=TeX,Scale=1}
\fi
\usepackage{lmodern}
\ifPDFTeX\else  
\fi
\IfFileExists{upquote.sty}{\usepackage{upquote}}{}
\IfFileExists{microtype.sty}{
  \usepackage[]{microtype}
  \UseMicrotypeSet[protrusion]{basicmath} 
}{}
\makeatletter
\@ifundefined{KOMAClassName}{
  \IfFileExists{parskip.sty}{%
  }{
    \setlength{\parindent}{0pt}
    \setlength{\parskip}{6pt plus 2pt minus 1pt}}
}{
  \KOMAoptions{parskip=half}}
\makeatother
\usepackage{xcolor}
\makeatletter
\ifx\paragraph\undefined\else
  \let\oldparagraph\paragraph
  \renewcommand{\paragraph}{
    \@ifstar
      \xxxParagraphStar
      \xxxParagraphNoStar
  }
  \newcommand{\xxxParagraphStar}[1]{\oldparagraph*{#1}\mbox{}}
  \newcommand{\xxxParagraphNoStar}[1]{\oldparagraph{#1}\mbox{}}
\fi
\ifx\subparagraph\undefined\else
  \let\oldsubparagraph\subparagraph
  \renewcommand{\subparagraph}{
    \@ifstar
      \xxxSubParagraphStar
      \xxxSubParagraphNoStar
  }
  \newcommand{\xxxSubParagraphStar}[1]{\oldsubparagraph*{#1}\mbox{}}
  \newcommand{\xxxSubParagraphNoStar}[1]{\oldsubparagraph{#1}\mbox{}}
\fi
\makeatother

\usepackage{longtable,booktabs,array}
\usepackage{calc} 
\usepackage{etoolbox}
\makeatletter
\patchcmd\longtable{\par}{\if@noskipsec\mbox{}\fi\par}{}{}
\makeatother
\IfFileExists{footnotehyper.sty}{\usepackage{footnotehyper}}{\usepackage{footnote}}
\makesavenoteenv{longtable}
\usepackage{graphicx}
\makeatletter
\def\maxwidth{\ifdim\Gin@nat@width>\linewidth\linewidth\else\Gin@nat@width\fi}
\def\maxheight{\ifdim\Gin@nat@height>\textheight\textheight\else\Gin@nat@height\fi}
\makeatother
\setkeys{Gin}{width=\maxwidth,height=\maxheight,keepaspectratio}
\makeatletter
\def\fps@figure{htbp}
\makeatother

\makeatletter
\@ifpackageloaded{caption}{}{\usepackage{caption}}
\AtBeginDocument{%
\ifdefined\contentsname
  \renewcommand*\contentsname{Table of contents}
\else
  \newcommand\contentsname{Table of contents}
\fi
\ifdefined\listfigurename
  \renewcommand*\listfigurename{List of Figures}
\else
  \newcommand\listfigurename{List of Figures}
\fi
\ifdefined\listtablename
  \renewcommand*\listtablename{List of Tables}
\else
  \newcommand\listtablename{List of Tables}
\fi
\ifdefined\figurename
  \renewcommand*\figurename{Figure}
\else
  \newcommand\figurename{Figure}
\fi
\ifdefined\tablename
  \renewcommand*\tablename{Table}
\else
  \newcommand\tablename{Table}
\fi
}
\@ifpackageloaded{float}{}{\usepackage{float}}
\floatstyle{ruled}
\@ifundefined{c@chapter}{\newfloat{codelisting}{h}{lop}}{\newfloat{codelisting}{h}{lop}[chapter]}
\floatname{codelisting}{Listing}

\makeatother
\makeatletter
\@ifpackageloaded{caption}{}{\usepackage{caption}}
\@ifpackageloaded{subcaption}{}{\usepackage{subcaption}}
\makeatother

\ifLuaTeX
  \usepackage{selnolig}  
\fi
\usepackage[]{natbib}
\usepackage{bookmark}

\IfFileExists{xurl.sty}{\usepackage{xurl}}{} 
\hypersetup{
  pdftitle={Title},
  pdfauthor={Author 1; Author 2},
  pdfkeywords={3 to 6 keywords, that do not appear in the title},
  colorlinks=true,
  linkcolor={blue},
  filecolor={Maroon},
  citecolor={Blue},
  urlcolor={Blue},
  pdfcreator={LaTeX via pandoc}}

\newcommand{\anon}{1}

\begin{document}

\def\spacingset#1{\renewcommand{\baselinestretch}%
{#1}\small\normalsize} \spacingset{1}


\if1\anon
{
  \title{\bf A novel finite-sample testing procedure for composite null hypotheses via pointwise rejection}
  \author{Joonha Park\thanks{
    This author gratefully acknowledges the General Research Fund of the College of Liberal Arts and Sciences at the University of Kansas and the Don and Pat Morrison Foundation.}\hspace{.2cm}\\
    Department of Mathematics, University of Kansas\\
    and \\
    Ming Wang \\
    Department of Mathematics, University of Kansas}
  \date{}
  \maketitle
} \fi

\if0\anon
{
  \bigskip
  \bigskip
  \bigskip
  \begin{center}
    {\LARGE\bf A novel finite-sample testing procedure for composite null hypotheses via pointwise rejection}
\end{center}
  \medskip
} \fi

\bigskip
\begin{abstract}
  We propose a novel finite-sample procedure for testing composite null hypotheses.
  Traditional likelihood ratio tests based on asymptotic $\chi^2$ approximations can exhibit substantial size distortions in small samples.
  Our procedure rejects the composite null hypothesis $H_0: \theta \in \Theta_0$ if the simple null hypothesis $H_{0,t}: \theta = \theta_t$ is rejected for every $\theta_t \in \Theta_0$, using a suitably inflated significance level.
  We derive a general formula for calibrating this inflated level based on the geometry of the null region $\Theta_0$.
  By employing finite-sample methods for testing the simple hypotheses $H_{0,t}: \theta = \theta_t$, the proposed pointwise testing procedure reduces the discrepancy between the actual and nominal significance levels, even in small samples.
  Whereas the traditional likelihood ratio test applies when the null region is defined solely by equality constraints, the proposed approach extends to null hypotheses defined by both equality and inequality constraints.
  We also discuss the application of the pointwise testing framework to composite null hypotheses expressed as unions of several component regions and to models involving nuisance parameters.
  Through several examples, we demonstrate numerically that the proposed test achieves accurate Type I error control in both small- and large-sample settings.
\end{abstract}

 
\noindent%
\vfill

\newpage
\spacingset{1.8} 

\section{Introduction}\label{sec:intro}

Consider a hypothesis test
\[
  H_0: \theta \in \Theta_0, \quad H_1: \theta \in \Theta_1,
\]
where $\theta$ denotes the parameter for the population distribution from which a sample is observed, and $\Theta_0$ and $\Theta_1$ are disjoint subsets of the parameter space $\Theta$.
A hypothesis is called \emph{simple} if it specifies $\theta$ to a single point, that is, if $\Theta_0$ is a singleton set.
A hypothesis is called \emph{composite} if it is not simple.

The likelihood ratio test (LRT) is widely used for testing both simple and composite null hypotheses \citep{vandervaart1998}.
If the null hypothesis is simple, $H_0: \theta = \theta_0$, the LRT rejects $H_0$ if and only if the likelihood ratio does not exceed a critical value $c$: 
\[
  \Lambda(\theta_0, \{\theta_0\}^\mathsf{c}; x) := \frac{L(\theta_0;x)}{\sup_{\theta \neq \theta_0} L(\theta;x)} \leq c,
\]
where $L(\theta;x)$ denotes the likelihood of $\theta$ given the observed sample $x$.
For a predetermined significance level $\alpha$, the critical value $c$ is the largest number satisfying
\[
  P_{\theta_0} ( \Lambda(\theta_0, \{\theta_0\}^\mathsf{c}; X) \leq c ) \leq \alpha,
\]
where $P_{\theta_0}$ indicates the probability distribution under $\theta_0$.
When the null hypothesis is composite, the likelihood ratio statistic is given by 
\[
  \Lambda(\Theta_0, \Theta_1; x) := \frac{\sup_{\theta \in \Theta_0} L(\theta;X)}{\sup_{\theta \in \Theta_1} L(\theta;X)},
\]
whose distribution generally depends on the parameter $\theta_0$ under which the sample $X$ is observed.
In this case, the significance level is defined as the supremum of the Type I error rate over all $\theta_0 \in \Theta_0$.
In many situations, however, finding the maximum value of $c$ satisfying 
\[
  \sup_{\theta_0 \in \Theta_0} P_{\theta_0} ( \Lambda(\Theta_0, \Theta_1; X) \leq c ) \leq \alpha
\]
is challenging, since the distribution of $\Lambda(\Theta_0, \Theta_1; X)$ is typically analytically intractable.

These challenges are often circumvented using large-sample approximations.
If $\Theta = \mathbb R^d$ and $\Theta_0$ is a $d - d_g$ dimensional subset defined by
\begin{equation}
  \Theta_0 = \{\theta \in \Theta: g(\theta) = 0 \},
  \label{eqn:Theta_0_eq}
\end{equation}
where $g:\Theta \to \mathbb R^{d_g}$ is a $d_g$ dimensional vector-valued function, then under suitable regularity conditions,
\[
P_{\theta_0}( \Lambda(\Theta_0, \Theta_1; X) \leq c) \to P[\chi^2_{d_g} \geq -2 \log c] \quad \text{for every } \theta_0 \in \Theta_0,
\]
as the sample size $n \to \infty$ \citep{wilks1938}.
This result greatly simplifies the construction of the critical region: 
\begin{equation}
  -2 \log \Lambda(\Theta_0, \Theta_1; X) \geq \chi_{1-\alpha, d_g}^2,
  \label{eqn:wilks}
\end{equation}
where $\chi_{1-\alpha, d_g}^2$ denotes the $1-\alpha$ quantile of the $\chi_{d_g}^2$ distribution.
However, this test based on a large-sample approximation may exhibit a substantial discrepancy between the nominal and actual significance levels for small sample sizes.

In this paper, we propose a novel approach for testing composite hypotheses, $H_0: \theta \in \Theta_0$, that reduces this discrepancy in finite-sample settings.
Intuitively, $H_0: \theta \in \Theta_0$ should not be rejected if there is at least one parameter value in $\Theta_0$ that explains the observed data reasonably well.
In other words, rejection of $H_0$ should indicate that every parameter value in $\Theta_0$ is inconsistent with the observed data.
This motivates a pointwise testing framework, in which $H_0: \theta \in \Theta_0$ is rejected if and only if $H_{0,t}: \theta = \theta_t$ is rejected for every $\theta_t \in \Theta_0$.
However, the probability of rejecting $H_{0,t}$ for all $\theta_t \in \Theta_0$ simultaneously is smaller than the probability of rejecting $H_{0,t}$ for a single $\theta_t$, making the resulting test conservative.
To address this issue, we require each pointwise test $H_{0,t}: \theta = \theta_t$ to be rejected at an inflated significance level $\alpha'$, which is greater than the target significance level $\alpha$.
The calibration of $\alpha'$, which depends on the geometry of $\Theta_0$, is discussed in Section~\ref{sec:calibration}.

The proposed pointwise testing procedure (PTP) can be understood as a union-intersection method, with the null hypothesis expressed as a union of singleton sets,
\[
  H_0: \theta \in \bigcup_{\theta_t \in \Theta_0} \{ \theta_t \}.
\]
The corresponding rejection region is therefore the intersection of the rejection regions for the pointwise hypotheses $H_{0,t}: \theta \in \{ \theta_t \}$ over $\theta_t \in \Theta_0$.
One of the major contributions of our work is the development of a general formula for determining the modified significance level $\alpha'$ for a given target significance level $\alpha$ (Equation~\ref{eqn:alpha_prime_general}).

There are two major advantages of the PTP.
First, the proposed procedure can achieve an actual size close to the nominal level $\alpha$ by employing finite-sample methods for pointwise testing of null hypotheses $H_{0,t}: \theta = \theta_t$, which are \emph{simple}.

Second, the PTP applies to a substantially broader class of null hypotheses than the traditional LRT.
The traditional LRT typically applies when $\Theta_0$ is a subset of $\Theta$ of the form \eqref{eqn:Theta_0_eq}, that is, when $H_0: \theta \in \Theta_0$ is specified solely by equality constraints.
By contrast, the PTP applies more generally to
\[
  \Theta_0 = \{ \theta \in \Theta: g(\theta) = 0, \, h(\theta) \leq 0 \},
\]
where $h: \Theta \to \mathbb R^{d_h}$ defines inequality constraints, with $h(\theta) \leq 0$ understood componentwise.

The class of testable null regions can be broadened even further by accommodating unions, such as
\[
  \Theta_0 = \bigcup_{i=1}^k \Theta_0^{(i)},
\]
where each $\Theta_0^{(i)}$ is specified by equality and inequality constraints and satisfies certain conditions.
The PTP also provides a principled approach to inference in the presence of nuisance parameters and allows for the construction of confidence regions for the parameter of interest.

The rest of this paper is organized as follows.
Section~\ref{sec:test} describes the new testing procedure and develops the formula for determining $\alpha'$.
Section~\ref{sec:confreg} applies the method to the case where the model contains nuisance parameters.
Section~\ref{sec:applications} illustrates our testing procedure for various composite null hypotheses and numerically demonstrates that its rejection probabilities remain relatively close to the nominal significance level, even when the sample size is small.
Notably, we apply our method to test the monotonicity of means of independent normal populations with \emph{unknown} population variances, extending previous work in the literature.
Section~\ref{sec:discussion} concludes with discussion.

\section{Method} \label{sec:test}
We propose a method for hypothesis test $H_0: \theta \in \Theta_0$ versus $H_1: \theta \notin \Theta_0$ where, given an observed sample $x=(x_1,\dots, x_n)$, the composite null hypothesis $H_0$ is rejected if and only if 
\[
  H_{0,t}: \theta = \theta_t \text{ is rejected in favor of }H_{1,t}: \theta\neq \theta_t \text{ at level } \alpha' \text{ for every }\theta_t \in \Theta_0.
\]
For any positive integer $n$, we will write $[n] := \{1, 2, \dots,
n\}$.
We assume that $\Theta = \mathbb R^d$ and consider null regions $\Theta_0$ defined by equality and inequality
constraints:
\begin{equation}
  \Theta_0 := \{ \theta \in \Theta: g_i(\theta) = 0 \text{ for } i\in
  [d_g] \text{ and } h_j(\theta) \leq 0 \text{ for } j\in [d_h]\}.
  \label{eqn:Theta_0}
\end{equation}

\subsection{Calibration of the Modified Significance Level $\alpha'$}\label{sec:calibration}
A traditional approach to testing a finite collection of component hypotheses
\[
  H_0: S_1 \text{ and } S_2 \text{ and } \dots \text{ and } S_B,
\]
where $H_0$ comprises $B$ statements, is to use the Bonferroni correction.
Under this approach, $H_0$ is rejected if at least one statement $S_b$, $b \in [B]$, is rejected at the Bonferroni-corrected significance level $\alpha^\text{BC} = \alpha/B$.
Although this approach ensures that the Type I error probability is bounded above by $\alpha$, it can be overly conservative \citep{perneger1998}.

Another approach to multiple hypothesis testing is to combine p-values of individual component hypotheses.
Test statistics such as (a) $\min(p_i)$ \citep{tippett1931methods}, (b) $-2\sum \log p_i$ \citep{fisher1932statistical}, and (c) $\sum \Phi^{-1}(1-p_i)$ \citep{liptak1958combination} can be used, whose null distributions are derived from the joint distribution of the individual p-values under the null hypotheses \citep{mudholkar1979logit}.

The PTP differs from conventional multiple-testing approaches in two important respects.
First, the PTP is based on expressing the null hypothesis as a union of pointwise hypotheses indexed by $\theta_t \in \Theta_0$, which may constitute a continuum.
Second, rather than applying a generic multiple-testing correction or combining p-values, the PTP calibrates the rejection level based on the geometry of the constraints defining $\Theta_0$.
Specifically, we develop a formula for determining $\alpha'$ such that
\[
  \sup_{\theta \in \Theta_0} P_\theta(H_{0,t}: \theta = \theta_t
  \text{ is rejected at level } \alpha' \text{ for every } \theta_t
  \in \Theta_0 ) \approx \alpha.
\]

\begin{figure}[t]
  \centering
  \includegraphics[width=0.8\textwidth]{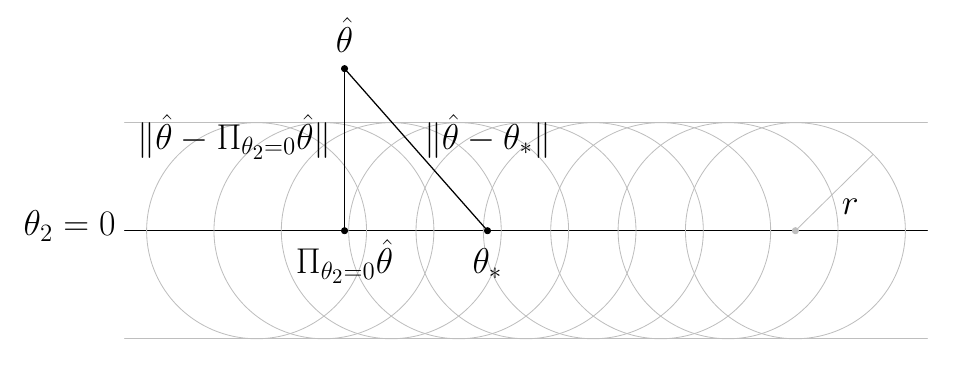}
  \caption{A diagram illustrating the pointwise testing procedure for $H_0: \theta_2 = 0$.}
  \label{fig:diagram_simple}    
\end{figure}

To illustrate the idea, consider a two dimensional model parameter $\theta = (\theta_1, \theta_2) \in \mathbb R^2$ and a null hypothesis $H_0: \theta_2 = 0$.
Suppose that the true parameter $\theta_*$ lies on the line $\{\theta: \theta_2 = 0\}$, and assume that the maximum likelihood estimator (MLE) follows the distribution
\[
  \hat \theta \sim \mathcal N(\theta_*, I_2),
\]
where $I_2$ denotes the two dimensional identity matrix.
Figure~\ref{fig:diagram_simple} illustrates the PTP.
Since the squared distance $\Vert \hat \theta - \theta_* \Vert^2$ follows a chi-squared distribution with two degrees of freedom, $\chi^2_2$, the simple null hypothesis $H_{0,*}: \theta = \theta_*$ is rejected at significance level $\alpha'$ if
\[
  \Vert \hat \theta - \theta_* \Vert > \left( \chi^2_{1-\alpha', 2} \right)^{1/2} =: r.
\]
Under the PTP, $H_0: \theta_2 = 0$ is rejected if $H_{0,t} : \theta = \theta_t$ is rejected at level $\alpha'$ for every $\theta_t \in \Theta_0$.
This occurs if and only if
\[
  \Vert \hat \theta - \Pi_{\theta_2=0} \hat \theta \Vert > r,
\]
where $\Pi_{\theta_2=0} \hat \theta$ denotes the orthogonal projection of $\hat \theta$ onto $\{\theta: \theta_2 = 0 \}$.
However, $\Vert \hat \theta - \Pi_{\theta_2=0} \hat \theta \Vert^2$ follows the $\chi^2_1$ distribution, and hence
\[
  P_{\theta_*}(H_0\text{ is rejected}) = P_{\theta_*}(\Vert \hat \theta - \Pi_{\theta_2=0} \hat \theta \Vert^2 > r^2) = 1 - F_1(\chi^2_{1-\alpha', 2})
\]
where $F_1$ denotes the cumulative distribution function (cdf) of $\chi^2_1$.
The modified significance level $\alpha'$ can then be calibrated by the equation
\begin{equation}
  1-\alpha = F_1(\chi^2_{1-\alpha', 2}),
  \label{eqn:alpha_prime_d2_dh0}
\end{equation}
where $\alpha$ is the target significance level.

Generalizing this idea, we calibrate $\alpha'$ for the null region given by \eqref{eqn:Theta_0}.
We assume that the MLE is approximately distributed as
\[
  \hat \theta \sim \mathcal N(\theta_*, \mathcal I(\theta_*)^{-1}),
\]
where
\[
  \mathcal I(\theta_*) := -\mathbb E_{\theta_*} \frac{\partial^2}{\partial
    \theta^2} \ell(\theta)
\]
is the Fisher information corresponding to the full data log-likelihood $\ell(\theta)$ evaluated at $\theta_*$.
The probability of rejecting $H_{0,t}: \theta = \theta_t$ under $\theta_*$ can then be approximated by
\begin{equation}
  P_{\theta_*}[ (\hat\theta - \theta_t) \mathcal I(\theta_t) (\hat \theta
  - \theta_t) > \chi^2_{1-\alpha', d} \text{ for every } \theta_t \in
  \Theta_0 ].
  \label{eqn:rejprob_approx}
\end{equation}
We assume that this probability is maximized when $\theta_*$ is one of
the extreme points of $\Theta_0$.
We call a point $\theta \in \Theta_0$ an extreme point of $\Theta_0$
if the set of inequality constraints that are binding at $\theta$, denoted by
\[
  S(\theta) = \{ j \in [d_h] : h_j(\theta) = 0 \},
\]
has at least as many elements as $S(\theta')$ for any other $\theta' \in \Theta_0$.
For example, if $\Theta_0$ is a triangle in $\Theta=\mathbb R^2$ specified by
three linear inequalities
\[
  h_1(\theta) \leq 0, \quad h_2(\theta) \leq 0, \quad h_3(\theta) \leq
  0,
\]
then the set of extreme points consists of its three vertices:
\[
  \{\theta: h_1(\theta) = h_2(\theta) = 0 \text{ or } h_2(\theta) =
  h_3(\theta) = 0 \text{ or } h_1(\theta) = h_3(\theta) = 0 \}.
\]
We will assume that the set of points in $\Theta_0$ at which exactly $k$ inequality constraints are binding, that is,
\[
  \{ \theta \in \Theta_0 : |S(\theta)| = k \},
\]
forms a $d - d_g - k$ dimensional face of $\Theta_0$ for $k \in \{0, \dots, \min(d_h, d-d_g)\}$, understanding that a $0$ dimensional face is a vertex and a negative dimensional face is an empty set.
Thus, if $d_g + d_h \geq d$, there will be $d_h \choose d-d_g$ vertices that are extreme points of $\Theta_0$.
In this case, the maximum of the rejection probability \eqref{eqn:rejprob_approx} as $\theta_*$ ranges over all extreme points of $\Theta_0$ will be calibrated to match the target significance level $\alpha$.

For a given extreme point $\theta_*$ of $\Theta_0$, we consider reparametrization
\[
  \gamma = \mathcal I(\theta_*)^{1/2}(\theta - \theta_*)
\]
so that approximately we have
\begin{equation}
  \hat \gamma \sim \mathcal N(0, I_d),
  \label{eqn:gamma_hat_normal}
\end{equation}
where $\mathcal I(\theta_*)^{1/2}$ denotes the square root matrix of $\mathcal I(\theta_*)$.
We consider linearized constraints defined by
\begin{equation}
  \begin{split}
    \tilde g_i(\gamma) &:= \{ \mathcal I(\theta_*)^{-1/2} \nabla g_i(\theta_*) \} \,\cdot\, \gamma = 0, \quad i \in [d_g],\\
    \tilde h_j(\gamma) &:= \{ \mathcal I(\theta_*)^{-1/2} \nabla h_j(\theta_*) \} \,\cdot\, \gamma \leq 0, \quad j \in S(\theta_*).
  \end{split}
  \label{eqn:linearized_constraints}
\end{equation}
Note that we only consider the inequality constraints for $j \in S(\theta_*)$, namely, those that are binding at $\theta_*$.
We define the corresponding linearized null region by
\begin{equation}
  \Gamma_0 := \{ \gamma: \tilde g_i(\gamma) = 0 \text{ for } i \in [d_g] \text{ and } \tilde h_i(\gamma) \leq 0 \text{ for } j \in S(\theta_*) \}.
  \label{eqn:Gamma0}
\end{equation}
Then $H_0: \theta \in \Theta_0$ fails to be rejected approximately when the MLE $\hat \gamma$ of $\gamma$ falls in 
\begin{equation}
  \mathcal A := \{ \gamma: \Vert \gamma - \gamma_t \Vert^2 \leq \chi^2_{1-\alpha', d} \text{ for some } \gamma_t \in \Gamma_0 \}.
  \label{eqn:mathcalA}
\end{equation}
Since $\hat\gamma$ is approximately normally distributed (Equation~\ref{eqn:gamma_hat_normal}), the probability of failing to reject $H_0$ can be approximated by 
\[ 
  P_{\theta_*}(\hat \gamma \in \mathcal A) \approx \Phi_d(\mathcal A),
\]
where $\Phi_d$ denotes the distribution function of $\mathcal N(0, I_d)$.

Let
\[
  \rho(\gamma) := \arg\min_{\gamma'\in\Gamma_0} \Vert \gamma - \gamma' \Vert
\]
denote the point in $\Gamma_0$ closest to $\gamma$.
Proposition~\ref{prop:alpha_prime_general} provides a formula for computing $\Phi_d(\mathcal A)$ by partitioning $\mathcal A$ into subsets according to the face of the polytope $\Gamma_0$ that is closest to each point:
\[
  \mathcal A = \bigsqcup_{V \subseteq S(\theta_*)} \mathcal A_V,
\]
where
\[
  \mathcal A_V = \{ \gamma: \Vert \gamma - \Gamma_0 \Vert < r, ~ \tilde h_j(\rho(\gamma)) = 0 \text{ for } j \in V, \text{ and } \tilde h_j(\rho(\gamma)) < 0 \text{ for } j \in S(\theta_*) \setminus V \}.
\]

\begin{prop}\label{prop:alpha_prime_general}
  The probability that a multivariate normal random variable with mean zero and variance-covariance matrix $I_d$ falls in the set $\mathcal A$ is given by
  \begin{multline}
    \Phi_d(\mathcal A) = \sum_{V\subseteq S(\theta_*)} F_{d_g+|V|}(\chi^2_{1-\alpha', d}) \\
    \cdot \overline\Omega_{|S(\theta_*)|-|V|}\left( \text{ConeSpan}\left\{\Pi_{\tilde g=0, \tilde h_{S(\theta_*)\setminus\{i\}}=0}(-\nabla \tilde h_i): i \in S(\theta_*)\setminus V\right\} \right) \\
    \cdot \overline\Omega_{|V|} \left( \text{ConeSpan}\left\{ \Pi_{\tilde g=0}(\nabla \tilde h_i): i \in V \right\} \right),
    \label{eqn:phi_d_A}
  \end{multline}
  where $F_\nu$ denotes the cumulative distribution function of the chi-squared distribution with $\nu$ degrees of freedom.
\end{prop}

In \eqref{eqn:phi_d_A}, the cone spanned by basis vectors $v_1, \dots, v_k$ is denoted by
\[
  \text{ConeSpan}\left\{ v_1, v_2, \dots, v_k \right\}
  := \{ \gamma : \gamma = \sum_{j=1}^k c_j v_j \text{ for some } c_1, \dots, c_k > 0 \}.
\]
Additionally, $\Pi_{\tilde g=0, \tilde h_{S(\theta_*)\setminus\{i\}}=0}(-\nabla \tilde h_i)$ denotes the projection of $-\nabla \tilde h_i$ onto the subspace
\[
  \left\{\gamma \in \mathbb R^d: \tilde g_i(\gamma) = 0 \text{ for } i \in [d_g] \text{ and } \tilde h_j(\gamma) = 0 \text{ for } j \in S(\theta_*) \setminus \{i\} \right\},
\]
and $\Pi_{\tilde g=0}(\nabla \tilde h_i)$ is defined similarly.
The expression $\overline \Omega_k(C)$ denotes the normalized solid angle of the $k$ dimensional cone $C$, namely, the solid angle of the cone divided by the total surface area of the generalized sphere $S^{k-1}$ embedded in $\mathbb R^k$.
Thus, $\overline \Omega_2(\text{ConeSpan}\{v_1, v_2\})$ is equal to the angle between $v_1$ and $v_2$ divided by $2\pi$, and $\overline \Omega_1\left(\text{ConeSpan}\left\{ v\right\} \right) = \frac{1}{2}$ for any nonzero vector $v$ because $S^0 = \{1, -1\}$.
For zero dimensional cones, we have $\Omega_0(\{0\}) = 1$.

The formula \eqref{eqn:phi_d_A} is illustrated using a relatively simple example in Section~\ref{app:illust_alpha_prime_general} of the online supplementary material, while the proof of Proposition~\ref{prop:alpha_prime_general} is provided in Section~\ref{app:proof}.

Proposition~\ref{prop:alpha_prime_general} suggests the following equation for determining $\alpha'$ for the PTP for an arbitrary number of equality and inequality constraints:
\begin{multline}
  1-\alpha
  = \sum_{V\subseteq S(\theta_*)} F_{d_g+|V|}(\chi^2_{1-\alpha', d}) \\
  \cdot \overline\Omega_{|S(\theta_*)|-|V|}\left( \text{ConeSpan}\left\{\Pi_{\tilde g=0, \tilde h_{S(\theta_*)\setminus\{i\}}=0}(-\nabla \tilde h_i): i \in S(\theta_*)\setminus V\right\} \right) \\
  \cdot \overline\Omega_{|V|} \left( \text{ConeSpan}\left\{ \Pi_{\tilde g=0}(\nabla \tilde h_i): i \in V \right\} \right).
  \label{eqn:alpha_prime_general}
\end{multline}

We discuss two special cases.
When $d_h = 0$, we have $S(\theta_*) = \varnothing$, and \eqref{eqn:alpha_prime_general} simplifies to
\begin{equation}
  1-\alpha = F_{d_g}(\chi^2_{1-\alpha', d}),
  \label{eqn:alpha_prime_0}
\end{equation}
which agrees with \eqref{eqn:alpha_prime_d2_dh0} when $d=2$ and $d_g=1$.
When $d_h = 1$, \eqref{eqn:alpha_prime_general} becomes
\begin{equation}
  1-\alpha = \frac{1}{2} F_{d_g}(\chi^2_{1-\alpha', d}) + \frac{1}{2} F_{d_g+1}(\chi^2_{1-\alpha', d}).
  \label{eqn:alpha_prime_1}
\end{equation}
For these two cases, the choice of $\alpha'$ does not depend on the Fisher information $\mathcal I(\theta_*)$.

The normalized solid angle of a cone spanned by a set of vectors can, in principle, be computed using the formula developed by \citet{aomoto1977analytic, ribando2006measuring}.
Further details are provided in Section~\ref{app:solid_angle} of the online supplementary material.
However, the computational cost of evaluating the formula for a $k$ dimensional cone scales exponentially in $k \choose 2$, which can be prohibitively high for $k \geq 5$.
Instead, in our numerical experiments, we estimated the solid angle using a Monte Carlo method based on the proportion of independent random vectors from $\mathcal N(0, I_k)$ that fall inside the cone.

As \eqref{eqn:alpha_prime_general} relies on a local linearization of the constraints expressed in \eqref{eqn:linearized_constraints}, the PTP may exhibit nonnegligible bias when these constraints have substantial curvature over the region in which the MLE $\hat \theta$ is likely to lie.
An approximate condition under which the probability of rejecting $H_0$ is close to the target significance level may be expressed as
\begin{equation}
  \frac{\lambda_{\max}\left(\mathcal I(\theta_*)^{-1/2} \nabla^2 f(\theta_*) \mathcal I(\theta_*)^{-1/2} \right)}{\Vert \mathcal I(\theta_*)^{-1/2} \nabla f(\theta_*) \Vert} \lesssim 1,
\label{eqn:req_cond}
\end{equation}
where $f$ ranges over all $g_i$ for $i \in [d_g]$ and all $h_j$ for $j \in [d_h]$, $\nabla^2$ denotes the Hessian matrix, and $\lambda_{\max}(\cdot)$ denotes the largest absolute eigenvalue.
Since $\mathcal I(\theta_*)$ is the Fisher information for the full data likelihood, in the case of independent and identically distributed ({i.i.d.})\ observations, the left hand side of \eqref{eqn:req_cond} scales as the inverse square root of the sample size.
Further details regarding \eqref{eqn:req_cond} are provided in Section~\ref{app:req_cond} of the online supplementary material.

\subsection{Pointwise Hypothesis Testing Procedure}
The PTP proceeds as follows.
First, the modified significance level $\alpha'$ is determined using \eqref{eqn:alpha_prime_general}, which simplifies to \eqref{eqn:alpha_prime_0} or \eqref{eqn:alpha_prime_1} if $d_h \leq 1$.
If $d_h \geq 2$, the set of extreme points of $\Theta_0$ is identified.
If $d_g + d_h < d$, this set will typically be a $(d - d_g - d_h)$ dimensional face of $\Theta_0$.
In this case, a point $\theta_*$ in that face is selected, such as one that maximizes the likelihood function, and the formula \eqref{eqn:alpha_prime_general} is evaluated with $S(\theta_*) = [d_h]$.
If $d_g + d_h \geq d$, the set of extreme points will typically consist of $d_h \choose d-d_g$ points, since at most $d-d_g$ inequality constraints can be binding simultaneously with the $d_g$ equality constraints.
The maximum of the resulting $\alpha'$ values across these extreme points, as determined by \eqref{eqn:alpha_prime_general}, is then used as the final value of $\alpha'$.

The next step is to determine the set of test points.
In principle, the simple null hypothesis $H_{0,t}: \theta = \theta_t$ needs to be tested for every $\theta_t \in \Theta_0$.
However, we propose to use a single test point given by
\[
  \theta_t = \arg\max_{\theta \in \Theta_0} \ell(\theta),
\]
as the corresponding proxy test $H_{0,t}: \theta = \theta_t$ is likely to produce a p-value that is close to the highest p-value among all pointwise tests of simple null hypotheses.
If the p-value for $H_{0,t}: \theta = \theta_t$ is less than $\alpha'$, we reject the original null hypothesis $H_0: \theta \in \Theta_0$, and otherwise we fail to reject $H_0$.

There is no restriction on the method used to test the simple null hypothesis $H_{0,t}$.
However, the PTP can leverage finite-sample methods to achieve high accuracy in small-sample settings.
When analytically tractable finite-sample methods are not available, $H_{0,t}$ may be tested using the finite-sample null distribution of a test statistic estimated from repeated simulations under $\theta_t$.

A p-value for the composite null hypothesis $H_0: \theta \in \Theta_0$ can also be evaluated using the equation \eqref{eqn:alpha_prime_general}.
If we substitute the p-value for the proxy test $H_{0,t}: \theta = \theta_t$ for $\alpha'$ in \eqref{eqn:alpha_prime_general}, then the resulting sum in the right hand side produces one minus the p-value for the original hypothesis $H_0: \theta \in \Theta_0$:
\begin{multline}
  1 - (\text{p-value for }H_0)
  = \sum_{V\subseteq S(\theta_*)} F_{d_g+|V|}(\chi^2_{1-(\text{p-value for $H_{0,t}$}), d}) \\
  \cdot \overline\Omega_{|S(\theta_*)|-|V|}\left( \text{ConeSpan}\left\{\Pi_{\tilde g=0, \tilde h_{S(\theta_*)\setminus\{i\}}=0}(-\nabla \tilde h_i): i \in S(\theta_*)\setminus V\right\} \right) \\
  \cdot \overline\Omega_{|V|} \left( \text{ConeSpan}\left\{ \Pi_{\tilde g=0}(\nabla h_i): i \in V \right\} \right).
  \label{eqn:converted_p_value}
\end{multline}
The converted p-value \eqref{eqn:converted_p_value} is less than $\alpha$ if and only if the pointwise p-value for $H_{0,t}$ is less than $\alpha'$.

\begin{algorithm}[tp]
\DontPrintSemicolon
\caption{Pointwise Testing Procedure for Composite Null Hypotheses}\label{alg:hypotest}
\SetKwInOut{Input}{Input}
\SetKwInOut{Output}{Output}
\Input{
  Observed sample $x = (x_1, \dots, x_n)$; 
  Log-likelihood function $\ell(\theta; x)$;
  Parameter space $\Theta$ of dimension $d$;
  Composite null hypothesis $H_0: \theta \in \Theta_0 = \{\theta \in \Theta: g_i(\theta) = 0 \text{ for } i \in [d_g] \text{ and } h_j(\theta) \leq 0 \text{ for } j \in [d_h] \}$; 
  Significance level $\alpha$
}
\Output{p-value for $H_0$ and decision to reject $H_0$ or not}
\If {$d_g + d_h < d$}{
  Let $\mathcal F = \{ \theta \in \Theta_0: h_j(\theta) = 0 \text{ for } j \in [d_h] \}$ be the face of $\Theta_0$ consisting of the extreme points\\
  Let $\mathcal E$ be the singleton set consisting of $\arg \max_{\theta \in \mathcal F} \ell(\theta; x)$ 
} \Else {
  Let $\mathcal E = \{ \theta \in \Theta_0: h_j(\theta) = 0 \text{ for } d-d_g \text{ indices } j \text{ in } [d_h]\}$
}
For each extreme point $\theta_* \in \mathcal E$, linearize the constraints via \eqref{eqn:linearized_constraints} and compute $\alpha'$ according to \eqref{eqn:alpha_prime_general}. If $d_g + d_h > d$, take the maximum of the resulting $\alpha'$ values for the final value of $\alpha'$\\
Determine the test point $\theta_t = \arg\max_{\theta \in \Theta_0} \ell(\theta; x)$ \\
Carry out a finite-sample test of simple null hypothesis $H_{0,t}: \theta = \theta_t$ \\
If $H_{0,t}$ is rejected at level $\alpha'$, reject $H_0$; otherwise, fail to reject $H_0$ \\
The p-value for $H_0$ can be evaluated using \eqref{eqn:converted_p_value}.
\end{algorithm}

An alternative method is to select a finite number of proxy test points $\theta_t^{(j)} \in \Theta_0$ for $j \in [m]$ that cover the relevant region of $\Theta_0$ sufficiently well to ensure that at least one $\theta_t^{(j)}$ is close to the constrained maximizer of the likelihood, $\arg\max_{\theta \in \Theta_0} \ell(\theta)$.
  Specifically, the gap between the two smallest approximate proxy p-values, 
\[
  F_d\left( -2\{ \ell(\theta_t^{(j)}) - \sup_{\theta \in \Theta} \ell(\theta) \} \right), \qquad j \in [m],
\]
should be much smaller than $\alpha'$.
Then $H_0: \theta \in \Theta_0$ is rejected if $H_{0,t(j)}: \theta = \theta_t^{(j)}$ is rejected at level $\alpha'$ for every $j \in [m]$.
This alternative method replaces the maximization of $\ell(\theta)$ over $\theta \in \Theta_0$ with testing at a finite set of points approximating the constrained maximizer, with more proxy test points generally providing a more accurate approximation at the cost of additional computation.

If $d_h = 0$ and a large-sample LRT based on the $\chi^2$ approximation is used to test the simple null hypotheses $H_{0,t}: \theta = \theta_t$, the PTP reduces to the traditional LRT, whose critical region is given by \eqref{eqn:wilks}.
By employing finite-sample methods to test the simple null hypotheses $H_{0,t}$, however, the PTP can reduce the discrepancy between the actual and target significance levels, even though the modified significance level $\alpha'$ is determined using a large-sample approximation.
Empirical results presented in Section~\ref{sec:applications} support this claim.
Moreover, the PTP substantially extends the scope of null hypotheses $H_0: \theta \in \Theta_0$ to encompass those involving both inequality and equality constraints.

\subsection{Null Hypothesis Linked by `Or'}
The PTP may be extended to test null hypotheses consisting of statements linked by `or'.
Suppose that the null hypothesis is given by
\begin{equation}
  H_0: \theta \in \Theta_0 = \Theta_0^{(1)} \cup \Theta_0^{(2)} \cup \cdots \cup \Theta_0^{(k)}
  \label{eqn:H0_or}
\end{equation}
where $\Theta_0^{(i)}$ for each $i \in [k]$ is specified by equality and inequality constraints.
For each $i\in [k]$, consider a PTP where $H_0^{(i)}:\theta \in \Theta_0^{(i)}$ is rejected at level $\alpha$ if $H_{0,t}: \theta = \theta_t$ is rejected for every $\theta_t \in \Theta_0^{(i)}$ at a modified significance level ${\alpha'}^{(i)}$.
Assume that each test of $H_0^{(i)}$ is unbiased, meaning that the power function is bounded below by $\alpha$.

We reject $H_0: \theta \in \Theta_0^{(1)} \cup \cdots \cup \Theta_0^{(k)}$ if for every $i \in [k]$ and every $\theta_t \in \Theta_0^{(i)}$, the simple null $H_{0,t}: \theta = \theta_t$ is rejected at level ${\alpha'}^{(i)}$.
This ensures that, for every $\theta \in \Theta_0$,
\begin{equation}
  \begin{split}
    P_\theta (H_0 \text{ is rejected})
    &= P_\theta (H_0^{(i)} \text{ is rejected for every } i \in [k]) \\
    &\leq \min_{i \in [k]} P_\theta (H_0^{(i)} \text{ is rejected}) 
    \approx \alpha,
  \end{split}
  \label{eqn:H0_or_guarantee}
\end{equation}
because the probability of rejecting $H_0^{(i)}$ under $\theta$ is approximately $\alpha$ if $\theta \in \Theta_0^{(i)}$ and is greater than $\alpha$ if $\theta \notin \Theta_0^{(i)}$ due to the assumption that the test of $H_0^{(i)}$ is unbiased.
Hence, the rejection probability is approximately upper bounded by $\alpha$.

Moreover, provided that there exists an $i\in [k]$ and an extreme point $\theta_*$ of $\Theta_0^{(i)}$ such that
\begin{equation}
  \sum_{j\neq i} P_{\theta_*} ( H_0^{(j)} \text{ fails to be rejected} ) \ll \alpha,
  \label{eqn:H0_or_exact_cond}
\end{equation}
we have
\begin{equation}
  P_{\theta_*}(H_0 \text{ is rejected})
  \geq P_{\theta_*}(H_0^{(i)} \text{ is rejected})
  - \sum_{j\neq i} P_{\theta_*}(H_0^{(j)} \text{ fails to be rejected})
  \approx \alpha.
  \label{eqn:H0_or_other}
\end{equation}
From \eqref{eqn:H0_or_guarantee} and \eqref{eqn:H0_or_other}, it follows that
\[
  \sup_{\theta\in\Theta_0} P_\theta(H_0 \text{ is rejected}) \approx \alpha,
\]
so the size of the test is approximately equal to $\alpha$.

For example, consider a parameter $\theta = (\theta_1, \theta_2)$ and a hypothesis
\[
  H_0: \theta_1 \leq 0 \text{ or } \theta_2 \leq 0.
\]
For each $i\in \{1,2\}$, consider applying the PTP where $H_0^{(i)}: \theta_i \leq 0$ is rejected if $H_{0,t}:\theta = \theta_t$ is rejected at level $\alpha'$ for every $\theta_t$ whose $i$-th component is nonpositive, with $\alpha'$ satisfying \eqref{eqn:alpha_prime_1}, namely,
\begin{equation}
  \frac{1}{2} F_1(\chi^2_{1-\alpha', 2}) = \frac{1}{2} - \alpha.
  \label{eqn:d2dh1}
\end{equation}
If we consider $\theta_* = (0, \theta_{2*})$ where $\theta_{2*}$ is large and positive such that
\[
  P_{\theta_*}(H_0^{(2)}: \theta_2 \leq 0 \text{ fails to be rejected})
\]
is negligibly small, then
\[
  P_{\theta_*}(H_0: \theta_1 \leq 0 \text{ or } \theta_2 \leq 0 \text{ is rejected})
  \approx P_{\theta_*}(H_0^{(1)}: \theta_1 \leq 0 \text{ is rejected}) \approx \alpha,
\]
ensuring that the significance level is close to $\alpha$.

If no extreme point satisfying \eqref{eqn:H0_or_exact_cond} exists, however, the PTP for null hypothesis of the form \eqref{eqn:H0_or} may be conservative.
For example, consider $\theta = (\theta_1, \theta_2)$ and
\[
  H_0: (\theta_1 \leq 1 \text{ and }\theta_2 \leq 1) \text{ or } (\theta_1 \geq -1 \text{ and } \theta_2 \geq -1).
\]
In this case, the extreme point $(1,1)$ of
\[
  \Theta_0^{(1)} = \{ \theta: \theta_1\leq 1 \text{ and } \theta_2\leq 1\}
\]
is in the interior of $\Theta_0^{(2)} = \{ \theta: \theta_1 \geq -1 \text{ and } \theta_2 \geq -1\}$, and vice versa.
Therefore,
\[
  P_{(1,1)}(H_0^{(2)}: \theta_1 \geq -1 \text{ and } \theta_2 \geq -1 \text{ fails to be rejected})
\]
is non-negligible.
Therefore,
\begin{multline*}
  P_{(1,1)}(H_0: \theta \in \Theta_0^{(1)} \cup \Theta_0^{(2)} \text{ is rejected})
  = P_{(1,1)}(H_0^{(1)}: \theta \in \Theta_0^{(1)} \text{ is rejected}) \\
  - P_{(1,1)}(H_0^{(1)}: \theta \in \Theta_0^{(1)} \text{ is rejected but } H_0^{(2)}: \theta \in \Theta_0^{(2)} \text{ fails to be rejected})
\end{multline*}
is significantly smaller than $\alpha$.

The test can be made less conservative in this example by obtaining $\alpha'$ from \eqref{eqn:alpha_prime_general} with $|S(\theta_*)|=1$ rather than $|S(\theta_*)|=2$, since at each boundary point of $\Theta_0$, only one inequality constraint defining $\Theta_0^{(i)}$ is binding for each $i \in \{1,2\}$.
Generalizing this ad hoc modification, however, is likely to be complicated, and we do not pursue it in this article.

\section{Inference in the Presence of Nuisance Parameters} \label{sec:confreg}

Suppose that the model parameter consists of two components, $\theta = (\psi, \phi) \in \Theta = \Psi \times \Phi$, where inference focuses on $\psi$.
The nuisance parameter $\phi$ often poses challenges for inference on $\psi$.
In this section, we demonstrate that our proposed procedure facilitates inference in the presence of nuisance parameters.

We consider a null hypothesis $H_0: \psi = \psi_0$ for some $\psi_0 \in \Psi$.
This null hypothesis can be expressed as $H_0: \theta \in \Theta_0$, where $\Theta_0 = \{ (\psi_0, \phi): \phi \in \Phi \}$.
If we write $\dim(\Psi) = d_\psi$ and $\dim(\Phi) = d_\phi$, the number of equality constraints $d_g$ is $d_\psi$, and the dimension of the parameter space $d$ is $d_\psi + d_\phi$.

Various approaches have been proposed to address the nuisance parameters in inference.
\citet{basu1977} discussed eliminating the nuisance parameter based on the generalized sufficiency or the generalized conditionality principle.
The generalized sufficiency principle states that, if there exists a statistic $Y = u(X)$ whose marginal distribution does not depend on $\phi$ and the conditional distribution of the data given $Y$ does not depend on $\psi$, then the inference on $\psi$ should be based on the observed value of $Y$.
This principle supports the restricted maximum likelihood (REML) method, where a test on $H_0: \psi = \psi_0$ is performed based on the statistic 
\[
  \Lambda^\text{R}(\psi_0, \Psi; x) = \frac{f_Y(u(x);\psi_0)}{\sup_{\psi \in \Psi} f_Y(u(x); \psi)},
\]
where $f_Y(y; \psi)$ denotes the probability density of $Y=y$ under $\psi$.
An alternative method is supported by the generalized conditionality principle that, if there exists a statistic $W = v(X)$ whose distribution does not depend on $\psi$ but the conditional distribution of the data given $W$ does not depend on $\phi$, then the inference on $\psi$ should be based on the conditional distribution of the data given the observed value of $W$.
Both of these approaches, however, rely on the existence of a statistic where either its marginal distribution or the distribution of the data given the statistic depends only on $\psi$.

Another approach to infer $\psi$ in the presence of a nuisance parameter $\phi$ is to consider the profile likelihood $L_p(\psi;x) = \sup_{\phi \in \Phi} \ell(\psi, \phi; x)$.
\citet{royall2000} justified inference based on the profile likelihood ratio by developing a large-sample approximation to the probability of observing misleading statistical evidence.
\citet{reid2003} proposed an expression for the p-value based on the adjusted profile likelihood derived from higher-order approximations.
These methods, however, provide little information regarding the bias in finite-sample settings.

Under the PTP, the null hypothesis $H_0: \psi = \psi_0$ is rejected at level $\alpha$ if $H_{0,t}: \theta = (\psi_0, \phi_t)$ is rejected in favor of $H_{1,t}: \theta \neq (\psi_0, \phi_t)$ at level $\alpha'$ for every $ \phi_t \in \Phi$.
Here, $\alpha'$ is determined based on \eqref{eqn:alpha_prime_0}:
\[
  \alpha' = 1 - F_{d_\psi+d_\phi}(\chi^2_{1-\alpha, d_\psi}).
\]
In practice, a single proxy test point $\phi_t = \arg \max_{\phi\in \Phi} \ell(\psi_0, \phi)$ can be used.
Alternatively, a finite number of proxy test points $\phi_t^{(j)}$, $j \in [m]$ may be selected in a neighborhood of the $\phi$-component of the MLE, $\hat \theta = (\hat\psi, \hat\phi)$, and reject $H_0: \psi = \psi_0$ if $H_{0,t(j)}: \theta = (\psi_0, \phi_t^{(j)})$ is rejected at level $\alpha'$ for every $j$.

The PTP leads to the construction of a confidence region for $\psi$ as follows:
\[
  \begin{split}
    \mathcal R
    &= \left\{\psi_0 \in \Psi : \, H_{0,t}:\theta = (\psi_0, \phi_t) \text{ is not rejected at level } \alpha' \text{ for some } \phi_t \in \Phi \right\} \\
    &= \bigcup_{\phi_t\in\Phi} \left\{ \psi_0 \in \Psi: \, H_{0,t}: \theta = (\psi_0, \phi_t) \text{ is not rejected at level } \alpha' \right\}.
  \end{split}
\]
This confidence region may be approximated by
\begin{equation}
  \mathcal R
  \approx \bigcup_{j=1}^m \left\{ \psi_0 \in \Psi: \, H_{0,t(j)}: \theta = (\psi_0, \phi_t^{(j)}) \text{ is not rejected at level } \alpha' \right\},
  \label{eqn:nuisance_conf_region}
\end{equation}
which expresses the $(1-\alpha)$-confidence region for $\psi$ as a union of $(1-\alpha')$-confidence regions for $\psi$, each corresponding to a fixed, known value $\phi = \phi_t^{(j)}$.


\section{Applications}\label{sec:applications}

\subsection{Testing a Null Hypothesis with $\theta \in [a,b]$}\label{sec:H0_interval}
We first demonstrate the concept of the PTP through a simple example.

\medskip
\noindent \textbf{Example~\ref{sec:H0_interval}.} 
Consider a normal random variable $X$ with mean $\mu$ and unknown variance $\sigma^2$ and null hypotheses of the form $H_0: \mu \geq \mu_0$ or $H_0: \mu \in [a,b]$. 

Suppose that a random sample $x_1, x_2, \dots, x_n$ is available.
For the composite null hypothesis $H_0: \mu \geq \mu_0$, the PTP is equivalent to the standard one-tailed $t$-test: $H_0$ is rejected if
\[
\frac{\bar x - \mu_0}{s/\sqrt{n}} < t_{1-\alpha, n-1},
\]
where $\bar x$ and $s$ denote the sample mean and the sample standard deviation.
To see this, note that Equation~\ref{eqn:alpha_prime_1} with $d=1$, $d_h=1$, and $d_g=0$ states
\[
  1-\alpha = \frac{1}{2} F_0(\chi^2_{1-\alpha',1}) + \frac{1}{2} F_1(\chi^2_{1-\alpha',1}) = \frac{1}{2} + \frac{1}{2}(1-\alpha'),
\]
and thus $\alpha' = 2\alpha$.
Thus, $H_0: \mu \geq \mu_0$ is rejected if
\[
\left| \frac{\bar x - \mu_t}{s/\sqrt{n}} \right| > t_{1-\frac{\alpha'}{2}, n-1} = t_{1-\alpha, n-1} \text{ for every } \mu_t \geq \mu_0,
\]
which occurs if and only if
\[
\frac{\bar x - \mu_0}{s/\sqrt{n}} < t_{1-\alpha, n-1},
\]
coinciding exactly with the one-tailed $t$-test.

The PTP can also be used to test null hypotheses of the form $H_0: \mu \in [a,b]$.
In this case, the null hypothesis is rejected if 
\[
\left | \frac{\bar x - \mu_t}{s/\sqrt{n}} \right| > t_{1-\alpha, n-1} \text{ for every } \mu_t \in [a,b].
\]
The significance level of this test is given by 
\[
\begin{split}
  &\sup_{\mu\in[a,b]} P_\mu\left( \bar X \notin \left[ a - t_{1-\alpha,n-1} \frac{S}{\sqrt n}, b + t_{1-\alpha,n-1} \frac{S}{\sqrt n} \right] \right)\\
  &= P_a\left( \bar X < a - t_{1-\alpha,n-1} \frac{S}{\sqrt n} \text{ or } \bar X > b + t_{1-\alpha,n-1} \frac{S}{\sqrt n} \right)\\
\end{split}
\]
which is approximately equal to $\alpha$ if $\sqrt n (b-a)/\sigma \gg 1$.

For comparison, consider the Bonferroni method for testing $H_0: \mu \in [a, b]$.
Since the null hypothesis can be expressed as 
\[
H_0: \mu \geq a \quad \text{and}\quad \mu \leq b,
\]
it is rejected if $H_0: \mu \geq a$ is rejected at a Bonferroni-corrected significance level $\alpha^\text{BC} := \alpha/2$, or if $H_0: \mu \leq b$ is rejected at level $\alpha^\text{BC}$.
The Bonferroni method ensures that 
\[
\sup_{\mu\geq a} P_\mu(H_0: \mu\geq a \text{ is rejected}) + \sup_{\mu\leq b} P_\mu(H_0: \mu\leq b \text{ is rejected}) \leq \alpha.
\]
However, since the suprema are attained at different endpoints, namely $\mu=a$ and $\mu=b$, their sum is substantially larger than
\[
\sup_{\mu \in[a,b]} P_\mu(H_0: \mu \geq a \text{ is rejected or } H_0: \mu \leq b \text{ is rejected} ),
\]
which is approximately equal to the actual size of the test:
\[
\sup_{\mu \in [a,b]} P_\mu(H_0: \mu \in [a,b] \text{ is rejected} ).
\]
Consequently, the Bonferroni method can be excessively conservative.
By contrast, our approach maintains the actual significance level close to the nominal level.

We numerically compared the PTP with the Bonferroni method for testing the composite null hypothesis $H_0: \mu \in [0,1]$.
We generated $10{,}000$ independent samples of size $n=20$ from $N(\mu, \sigma^2)$ with $\sigma=1$.
Under the PTP, $H_0$ is rejected if
\[
  \bar x \notin \Big[- t_{1-\alpha,n-1}\tfrac{S}{\sqrt n},\; 1 + t_{1-\alpha,n-1}\tfrac{S}{\sqrt n}\Big].
\]
For the Bonferroni test, $H_0$ is rejected if either of the one-sided hypotheses $H_0: \mu \geq a$ or $H_0: \mu \leq b$ is rejected at the Bonferroni-corrected level $\alpha/2$.

\begin{figure}[t]
    \centering
    \includegraphics[width=0.8\textwidth]{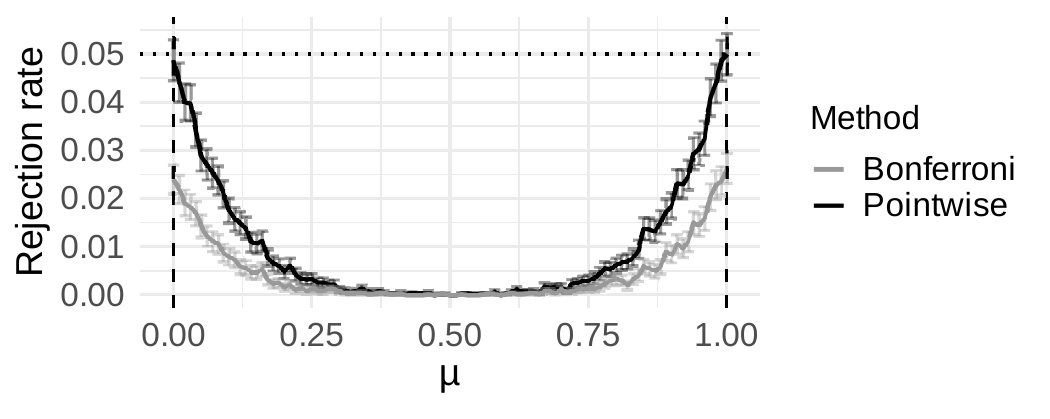}
    \caption{Empirical rejection rates of $H_0: \mu \in [0,1]$ for the PTP and the Bonferroni method as a function of the true mean $\mu$.}
    \label{fig:interval}
\end{figure}

Figure~\ref{fig:interval} displays the empirical rejection rate for $n=20$ as a function of the true mean $\mu$.
The maximum rejection rate within the null region $\mu \in [0,1]$ provides an estimate of the actual test size.
Our method achieves a maximum rejection rate near the nominal level $\alpha = 0.05$ at the boundary points ($\mu = 0$ and $\mu = 1$).
By contrast, the Bonferroni method attains a maximum rejection rate of about $0.025$, indicating that it is overly conservative.

\begin{figure}[t]
    \centering
    \includegraphics[width=0.8\textwidth]{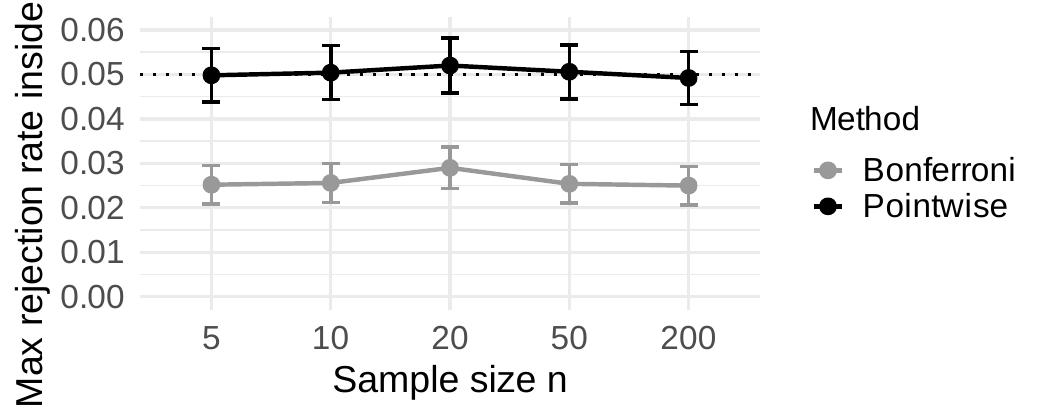}
    \caption{Empirical significance levels of the PTP and the Bonferroni method for varied sample sizes $n$.
    }
    \label{fig:interval_varied_n}
\end{figure}

We repeated the simulation study for sample sizes $n \in \{5, 10, 20, 50, 200\}$.
Figure~\ref{fig:interval_varied_n} shows that our method maintains accuracy across both small and large sample sizes, whereas the Bonferroni method yields significance levels roughly half the nominal level throughout.

\subsection{Testing for Two Parameters Linked by `Or'}\label{sec:H0_or}
\noindent \textbf{Example~\ref{sec:H0_or}.} 
Consider a linear model
\[
Y_i = \beta_0 + \beta_1 x_{i,1} + \beta_2 x_{i,2} + \epsilon_i,
\]
where the random errors $\epsilon_i$ are {i.i.d.}\ draws from $\mathcal N(0, \sigma^2)$.
We demonstrate testing of a null hypothesis
\[
  H_0: \beta_1 \leq 0 \text{ or } \beta_2 \leq 0
\]
using the PTP.
First, we find the ordinary least squares (OLS) estimates $(\hat \beta_0, \hat \beta_1, \hat \beta_2)$.
If $\hat \beta_1 \leq 0$ or $\hat \beta_2 \leq 0$, we fail to reject $H_0$.
If $\hat \beta_1 > 0$ and $\hat \beta_2 > 0$, we select $m'$ values $\beta_{1,t}^{(j)}$, $j \in [m']$, close to $\hat \beta_1$ and another set of $m'$ values $\beta_{2,t}^{(j)}$, $j \in \{m'+1, \dots, 2m'\}$, close to $\hat\beta_2$. 
Then, we conduct $m=2m'$ finite-sample $F$-tests for null hypotheses $H_{0,t(j)}: (\beta_1, \beta_2) = (\beta_{1,t}^{(j)}, 0)$ for $j\in [m']$ and $H_{0,t(j)}: (\beta_1, \beta_2) = (0, \beta_{2,t}^{(j)})$ for $j \in \{m'+1, \dots, 2m'\}$.
If the p-values for all of these $m$ proxy tests are less than $\alpha'$ determined by \eqref{eqn:d2dh1}, then we reject $H_0: \beta_1 \leq 0 \text{ or } \beta_2 \leq 0$, and otherwise, we fail to reject $H_0$.

\begin{table}[t]
\centering
\begin{tabular}{llllll}
\toprule
\multirow{2}{*}{$m$} & \multicolumn{5}{c}{Sample Size} \\ \cmidrule{2-6}
 & 5 & 10 & 20 & 50 & 100 \\ \midrule
10 & 5.32 (0.44) & 5.28 (0.44) & 5.52 (0.45) & 5.41 (0.44) & 5.09 (0.43) \\	
100 & 4.98 (0.43) & 4.68 (0.41) &	5.07 (0.43) & 5.41 (0.44) & 5.01 (0.43) \\
\bottomrule
\end{tabular}
\caption{Rejection rates (in \%) for varying sample sizes and numbers of proxy test points ($m$) under $H_0: \beta_1 \leq 0 \text{ or } \beta_2 \leq 0$ when the true parameter is $(\beta_1, \beta_2) = (1,0)$ in Example~\ref{sec:H0_or}. 
Parentheses indicate $1.96$ times the standard error.
}
\label{tab:H0_or}
\end{table}

We generated data of size $n \in \{5, 10, 20, 50, 100\}$, where $(x_{i,1}, x_{i,2})$ were independently drawn from a bivariate normal distribution with covariance matrix $I_2$, and parameters $\beta_0 = 0$ and $\sigma = 1$. 
When both OLS estimates $\hat \beta_1$ and $\hat \beta_2$ were positive, proxy test points $\hat \beta_{k,t}^{(j)}$ were chosen to be equally spaced in the interval $(0, 2\hat \beta_k)$ for $k=1,2$. 
Each experiment was repeated $10^4$ times.

Table~\ref{tab:H0_or} reports the proportion of rejections of $H_0: \beta_1 \leq 0 \text{ or } \beta_2 \leq 0$ at the nominal significance level $\alpha = 0.05$, with data generated under $(\beta_1, \beta_2) = (1,0)$. 
We considered $m=10$ and $m=100$ proxy test points. 
Across all sample sizes, the observed rejection rates were close to the nominal 5\% level, and results were similar for $m=10$ and $m=100$.

\begin{figure}[t]
\centering
\includegraphics[width=0.8\textwidth]{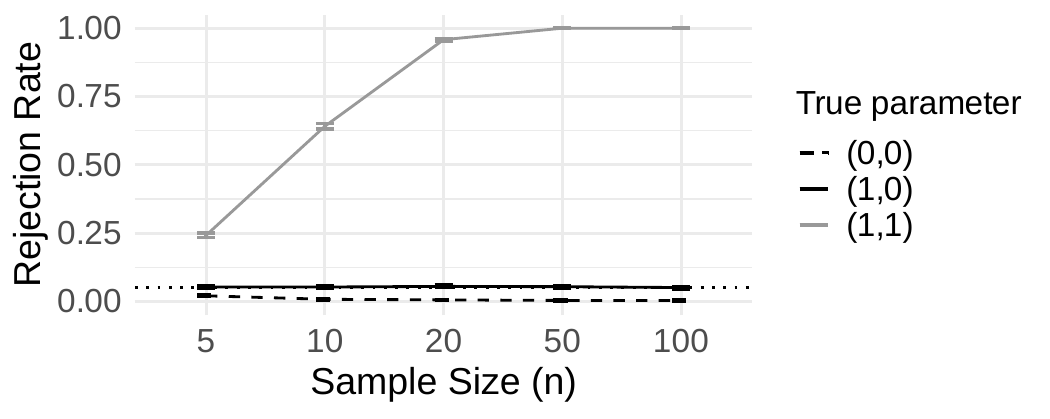}
\caption{
Empirical rejection probabilities of $H_0: \beta_1 \leq 0 \text{ or } \beta_2 \leq 0$ for varying sample sizes when the true parameter values are $(\beta_1, \beta_2) = (0,0)$, $(1,0)$, and $(1,1)$. 
Error bars represent $1.96$ times the standard errors, and the horizontal dotted line denotes the nominal significance level of 5\%.
}
\label{fig:H0_or_power}
\end{figure}

Figure~\ref{fig:H0_or_power} displays the empirical rejection probabilities of $H_0$ under different parameter settings.
When $(\beta_1, \beta_2) = (1,0)$, which lies on the boundary of $\Theta_0$, the Type I error rate is close to the nominal 5\% significance level, consistent with Table~\ref{tab:H0_or}.
When $(\beta_1, \beta_2) = (0,0)$, however, the rejection probability consistently falls below 0.05.
This is because the geometry of $\Theta_0$ in a neighborhood of $(0,0)$ is non-convex.
By contrast, when $(\beta_1, \beta_2) = (1,1)$, which lies outside $\Theta_0$, the rejection probability exceeds the nominal level, and the power increases with sample size.

\subsection{Testing in the Presence of a Nuisance Parameter}\label{sec:H0_nuisance}
Consider a parameter $\theta = (\psi, \phi)$, where $\psi$ denotes the parameter of interest and $\phi$ the nuisance parameter.
The PTP is applicable even when the likelihood ratio statistic $\sup_{\phi} L(\psi_0, \phi; X)/\sup_{\psi, \phi} L(\psi, \phi; X)$ or its distribution for given $\psi$ is not analytically tractable, as demonstrated by the following example.

\medskip
\noindent \textbf{Example~\ref{sec:H0_nuisance}.} 
Consider a simple linear regression
\begin{equation}
  Y_i = \psi \phi x_i + \psi \phi^2 + \epsilon_i, \quad i=1,\dots, n
\label{eqn:linmodel_nuisance}
\end{equation}
where $\epsilon_i$'s are {i.i.d.}\ Gaussian noise with mean zero and variance $\sigma^2$.
The focus is on testing $H_0: \psi = \psi_0$.
Denoting the OLS estimates for the intercept and the slope in \eqref{eqn:linmodel_nuisance} by $\hat \beta_0$ and $\hat \beta_1$, we obtain the MLE's for $\psi$ and $\phi$ given by $\hat{\phi} = \hat\beta_0/\hat\beta_1$ and $\hat{\psi}= \hat \beta_1/\hat{\phi}$. 
We then compute the residual sums of squares (RSS)
\[
\text{RSS}_\text{alt} = \sum_i (y_i - \hat y_{i, \text{alt}})^2, \quad
\text{RSS}_\text{null}(\psi_0, \phi_t) = \sum_i (y_i - \hat y_{i, \text{null}})^2
\]
where $\hat y_{i, \text{alt}} = \hat\psi \hat\phi x_i + \hat\psi \hat\phi^2$ and $\hat y_{i, \text{null}} = \psi_0 \phi_t x_i + \psi_0 \phi_t^2$.
The $F$-test statistic is given by 
\begin{equation}
F(\psi_0, \phi_t) = \frac{(\text{RSS}_{\text{null}}(\psi_0, \phi_t) - \text{RSS}_{\text{alt}})/2}{\text{RSS}_{\text{alt}}/(n - 2)},
\label{eqn:nuisance_F}
\end{equation}
and the p-value by $P[F_{2, n-2} > F(\psi_0, \phi_t)]$.

\begin{figure}[t]
  \centering
  \includegraphics[width=\textwidth]{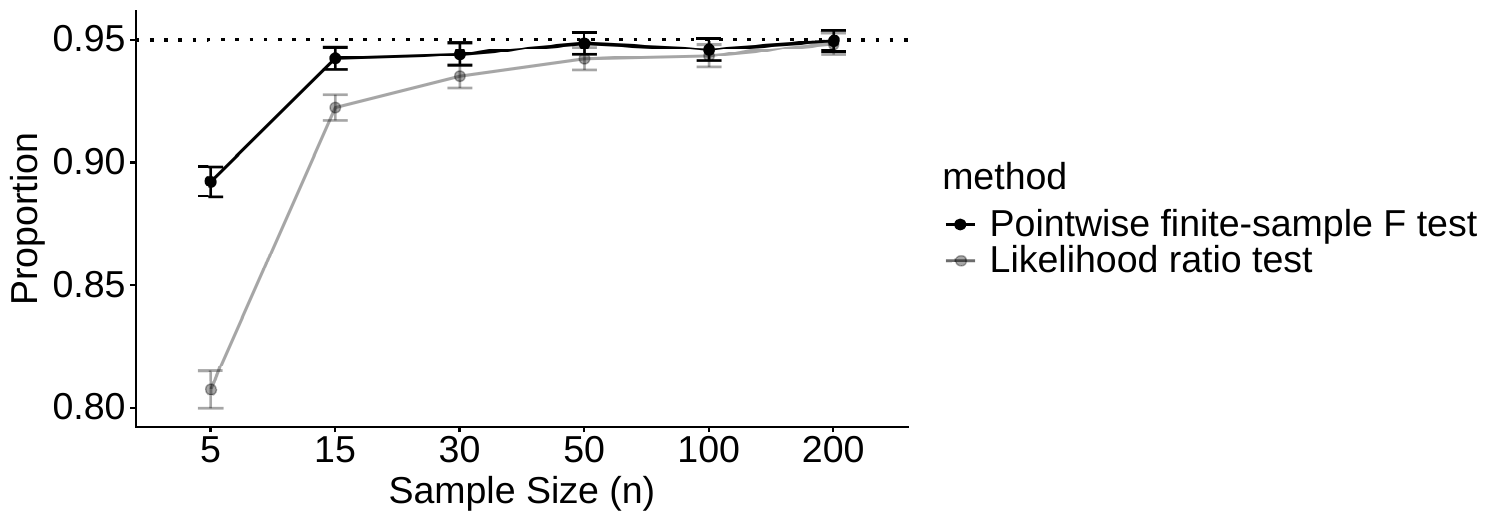}
  \caption{
  Proportion of constructed 95\% confidence intervals that contain the true parameter value $\psi = 1$ for Example~\ref{sec:H0_nuisance}, comparing the PTP using finite-sample $F$ test to the large-sample likelihood ratio test for varying sample sizes.
  }
  \label{fig:nuisance_F_chi}
\end{figure}

We numerically compared the accuracy of our proposed method with that of the LRT based on a large-sample $\chi^2$ approximation.
Samples of size $n \in \{5, 15, 30, 50, 100, 200\}$ were generated according to the model \eqref{eqn:linmodel_nuisance} with parameter values $\psi=1$, $\phi=2$, and error variance $\sigma^2=1$.
For each sample, we constructed 95\% confidence intervals for $\psi$ using \eqref{eqn:nuisance_conf_region}.
We selected proxy points $\phi_t^{(j)}$, $j\in [50]$, uniformly spaced in the interval $(\hat\phi - 5n^{-1/2}, \hat\phi + 5n^{-1/2})$, reflecting that the standard error of the MLE $\hat\phi$ scales as $n^{-1/2}$. 
For each $\phi_t^{(j)}$, an interval
\[
\mathcal R_j = \{ \psi_0 : F(\psi_0, \phi_t^{(j)}) \leq F_{2, n-2}(1-\alpha') \}
\]
was obtained, where $\alpha' = 1 - F_2(\chi^2_{0.95,1}) \approx 0.1465$ was determined according to \eqref{eqn:alpha_prime_0}, and $F_{2,n-2}(1-\alpha')$ denotes the $(1-\alpha')$-th quantile of the $F_{2,n-2}$ distribution.
For each sample size $n \in \{5, 15, 30, 50, 100, 200\}$, we repeated this procedure $10^4$ times.

For comparison, we also constructed 95\% confidence intervals for $\psi$ using the traditional LRT; further details are provided in Section~\ref{app:example_nuisance} of the online supplementary material.
For each sample size, $10^4$ confidence intervals were constructed.
Figure~\ref{fig:nuisance_F_chi} shows the proportion $\hat p$ of constructed confidence intervals containing the true parameter value $\psi=1$ for the PTP based on the finite-sample $F$ test and for the traditional LRT.
Error bars represent the margin of error, calculated as $1.96 \times \sqrt{\hat p (1-\hat p)/n}$.
As the sample size increases, the coverage probability of both methods approaches the nominal 95\% level.
For smaller sample sizes ($n < 100$), our method provides coverage that is closer to the nominal level than that of the traditional LRT.
Remarkably, the coverage bias remains modest even when $n=5$ or $15$.
Additional numerical results comparing the power of the PTP and the traditional LRT are provided in Section~\ref{app:example_nuisance} of the supplementary material.

\subsection{Testing Null Regions Defined by Equality and Inequality Constraints} \label{sec:H0_disc}
We consider an example where the null hypothesis $H_0: \theta \in \Theta_0$ is defined by both equality and inequality constraints.
In this case, $\Theta_0$ forms a lower dimensional manifold with boundary, and the traditional LRT is not applicable.

\noindent \textbf{Example~\ref{sec:H0_disc}.}
Suppose that the observed data are independently sampled from a multivariate normal distribution: 
\[
Y_i \overset{i.i.d.}{\sim} \mathcal{N}(\theta, \sigma^2 I_d), \quad i \in [n],
\]
where \( \theta \in \mathbb{R}^d \) is an unknown mean vector and $\sigma^2$ is known.
We  consider 
\[
H_0: \theta \in \Theta_0 \quad \text{vs.} \quad H_1: \theta \notin \Theta_0,
\]
where \( \Theta_0 \) is defined as the intersection of a $(d-d_g)$ dimensional subspace and a unit ball:
\[
\Theta_0 = \left\{ \theta \in \mathbb R^d: \, \theta_1 = \cdots = \theta_{d_g} = 0, \, \sum_{i=d_g+1}^{d} \theta_i^2 \leq 1 \right\}
\]
We apply the PTP using a single proxy test point, given by
\begin{equation}
\theta_t = \arg \max_{\theta \in \Theta_0} \ell(\theta; Y) = \arg \min_{\theta \in \Theta_0} \Vert \theta - \bar Y \Vert,
\label{eqn:disc_proj}
\end{equation}
where $\bar Y = \tfrac{1}{n} \sum_{i=1}^n Y_i \in \mathbb R^d$ is the unconstrained MLE. 

We compared the PTP with the \emph{universal inference} approach proposed by \citet{wasserman2020}.
The universal inference framework is broadly applicable to models that may not satisfy the regularity conditions required for the traditional LRT.
Under this framework, the data are divided into two independent parts, $Y^{(1)}$ and $Y^{(2)}$, and the parameter is estimated separately using each split.
For each $\theta_0 \in \Theta_0$, two held-out likelihood ratios are defined as
\[
U_1(\theta_0) = \exp\!\left\{
\ell(\hat{\theta}^{(2)}; Y^{(1)}) - \ell(\theta_0; Y^{(1)})
\right\},
\quad
U_2(\theta_0) = \exp\!\left\{
\ell(\hat{\theta}^{(1)}; Y^{(2)}) - \ell(\theta_0; Y^{(2)})
\right\},
\]
where $\hat \theta^{(1)}$ and $\hat \theta^{(2)}$ denote the parameter estimates based on the two data splits.
\citet{wasserman2020} proposed two testing procedures for $H_0: \theta \in \Theta_0$.
The \emph{split LRT} rejects $H_0$ if $U_1(\theta_0)> 1/\alpha$ for every $\theta_0 \in \Theta_0$, and the \emph{cross-fit LRT} rejects $H_0$ if $\{U_1(\theta_0) + U_2(\theta_0) \}/2 > 1/\alpha$ for every $\theta_0 \in \Theta_0$, where $\alpha$ is the nominal significance level.
In Example~\ref{sec:H0_disc}, we rejected $H_0$ when the corresponding test statistics, $U_1(\theta_t)$ or $\{U_1(\theta_t) + U_2(\theta_t)\}/2$, exceeded $1/\alpha$ for the single test value $\theta_t$ given by \eqref{eqn:disc_proj}.

\begin{table}[t]
\centering
\begin{tabular}{llllll}
\toprule
\multirow{2}{*}{method} & \multicolumn{5}{c}{Sample Size} \\ \cmidrule{2-6}
 & 5 & 10 & 30 & 100 & 1000 \\ \midrule
PTP & 6.81 (0.25) & 6.27 (0.24) & 5.99 (0.23) & 5.35 (0.22) & 5.25 (0.22) \\
split LRT & 0.08 (0.03) & 0.12 (0.03) & 0.06 (0.02) & 0.07 (0.03) & 0.05 (0.02) \\
cross-fit LRT & 0.07 (0.03) & 0.06 (0.02) & 0.04 (0.02) & 0.06 (0.02) & 0.04 (0.02) \\
\bottomrule
\end{tabular}
\caption{Estimated Type I error rates (in \%) for the PTP, the split LRT, and the cross-fit LRT applied to Example~\ref{sec:H0_disc}. Parentheses indicate 1.96 times the standard error.}
\label{tab:H0_disc}
\end{table}

We considered an example with $d=5$, $d_g=2$, and $\sigma=1$, for which $\alpha'$ for the PTP determined by Equation~\ref{eqn:alpha_prime_1} was approximately 0.2173 for the target significance level of $\alpha = 0.05$.
We conducted $4 \times 10^4$ simulations in which the PTP as well as the two variants of the split-sample test methods were applied for each sample size $n \in \{5, 10, 30, 100, 1000\}$.
In each simulation, data were generated from a multivariate normal distribution with mean $\theta_{\text{true}} = (0, 0, 0, 0, 1)$, which lies on the boundary of $\Theta_0$.
Table~\ref{tab:H0_disc} reports the estimated probability of rejecting $H_0: \theta \in \Theta_0$ at a nominal 5\% significance level.
As the sample size increased, the Type I error rate of the PTP converged to the nominal 5\% level, with an estimated bias of less than 2\% even at $n=5$.
In contrast, both methods proposed by \citet{wasserman2020} exhibited extremely low rejection rates---below 0.1\%---across all sample sizes. 
Although the universal inference framework guarantees Type I error control for almost any model through the Markov inequality, it can be highly conservative because the bound tends to be loose.
Additional numerical results on the statistical power of the methods are provided in Section~\ref{app:comp_univ} of the online supplementary material.

\begin{figure}[t]
    \centering
    \includegraphics[width=\textwidth]{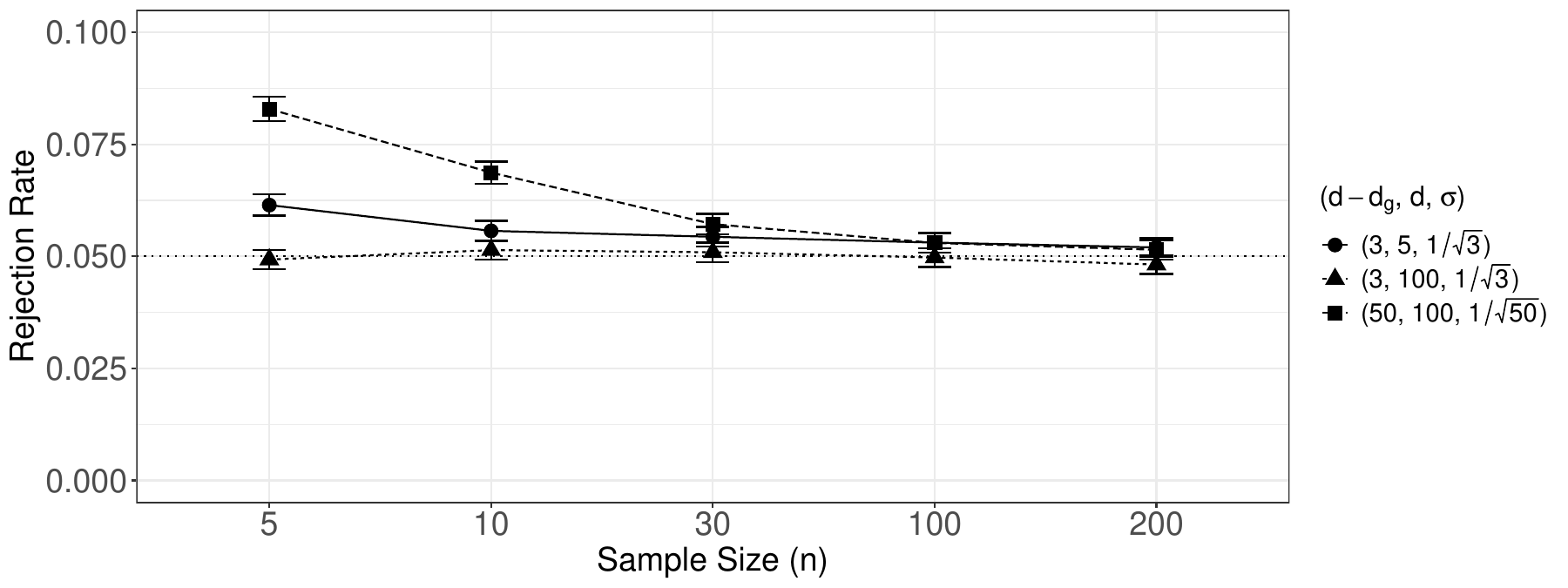}
    \caption{
      Empirical probabilities of rejecting $H_0: \theta \in \Theta_0$ for Example~\ref{sec:H0_disc} under varied dimensions and sample sizes. 
    Error bars represent $\pm 1.96$ times the standard error, and the horizontal dashed line shows the nominal 5\% significance level.
    }
    \label{fig:disc_extended}
\end{figure}

We also empirically evaluated the rejection probabilities of the PTP in higher dimensions.
We considered $(d-d_g, d) \in \{(3,5), (3, 100), (50, 100)\}$.
Since the inequality constraint $\sum_{i=d_g+1}^d \theta_i^2 \leq 1$ has a boundary with radius of curvature equal to 1, its linearization used in determining $\alpha'$ can introduce a substantial bias when
\[
  \left( \mathbb E \left\Vert \Pi_{g=0} (\bar Y) - \theta \right\Vert^2 \right)^{1/2} = \frac{\sigma}{\sqrt n} \sqrt{d-d_g} \gtrsim 1,
\]
where $\theta$ denotes the true parameter and $\Pi_{g=0}(\bar Y)$ the projection of $\bar Y$ onto the subspace $\{\theta: \theta_1 = \dots = \theta_{d_g} = 0\}$.
Thus we considered $\sigma = 1/\sqrt{d-d_g}$ in each of the three settings.
The sample size varied over $n \in \{5, 10, 30, 100, 200 \}$.
Figure~\ref{fig:disc_extended} shows the empirical rejection probabilities for $H_0: \theta \in \Theta_0$ at significance level $\alpha = 0.05$, based on $40{,}000$ replications under the true parameter $\theta$ satisfying $\theta_j = 0$ for $j \in [d-1]$ and $\theta_d=1$.
In all three settings, the empirical size is close to $\alpha = 0.05$, indicating that the PTP maintains reasonable accuracy even for small sample sizes in both low- and high-dimensional settings.

\subsection{Testing Monotonicity of Means}\label{sec:mean_monotone}
\noindent \textbf{Example~\ref{sec:mean_monotone}.}
We consider testing the monotonicity of means of independent normal populations.
Suppose that the $i$-th population follows the normal distribution with unknown mean $\mu_i$ and unknown variance $\sigma_i^2$ for $i \in [k]$.
We consider a null hypothesis
\[
  H_0: \mu_1 \leq \mu_2 \leq \cdots \leq \mu_k.
\]

\citet{mudholkar1989class} considered a closely related problem of testing equality of order-constrained means of normal populations with known and equal variances.
They constructed independent tests that, collectively, are equivalent to the test of equality of means under the order constraints, and used test statistics that combine the individual p-values in a multiple-testing framework \citep{mudholkar1979logit}.

The test of equality of order-constrained means can be reformulated so that there are no restrictions on the parameter space.
For instance, consider testing $H_0: \mu_1 = \cdots = \mu_k$ under the order constraints $\mu_1 \geq \mu_2 \geq \cdots \geq \mu_k$.
Under these constraints, the alternative hypothesis is that $\mu_i > \mu_{i+1}$ for at least one $i \in [k-1]$.
Thus, the corresponding critical region is equivalent to that for testing
\[
  H_0: \mu_1 \leq \mu_2 \leq \cdots \leq \mu_k
\]
without any restrictions on the parameter space, with $(\mu_1, \dots, \mu_k)$ is $\mathbb R^k$.
In this section, we consider the latter formulation, but in a more general setting, where the population variances $\sigma_i^2$ are assumed to be unequal and unknown.

The parameter vector $\theta = (\mu_1, \dots, \mu_k, \sigma_1, \dots, \sigma_k)$ is $2k$ dimensional, and the null hypothesis consists of $k-1$ linear inequality constraints, given by
\[
  h_i(\theta) = \mu_i - \mu_{i+1} \leq 0 \quad \text{for} \quad i \in [k-1],
\]
and zero equality constraints.
For the $i$-th population, $n_i$ independent observations $x_{ij}$, $j \in [n_i]$, are available with sample mean $\bar x_i$ and sample standard deviation $s_i$.

The profile log-likelihood is given by
\begin{equation}
  \sup_{\sigma_1, \dots, \sigma_k} \ell(\theta)
  = -\sum_{i=1}^k n_i \log \sum_{j=1}^{n_i}(x_{ij}-\mu_i)^2 + \text{const.}
  \label{eqn:mean_monotone_profile}
\end{equation}
Since extreme points of the null region satisfy $\mu_1 = \cdots = \mu_k$, we estimate the most likely extreme point by
\[
  \theta_* = (\mu_*, \dots, \mu_*, \sigma_{1*}, \dots, \sigma_{k*}),
\]
where
\[
  \mu_* = \arg \min_\mu \sum_{i=1}^k n_i \log \sum_{j=1}^{n_i} (x_{ij}-\mu)^2
\]
is numerically found and the corresponding variances are obtained by
\[
  \sigma_{i*}^2 = \frac{1}{n_i} \sum_{j=1}^{n_i} (x_{ij} - \mu_*)^2, \quad i\in [k].
\]
The modified significance level $\alpha'$ for the PTP is then determined using \eqref{eqn:alpha_prime_general}, where the Fisher information evaluated at $\theta_*$ is given by
\[
  \mathcal I(\theta_*) = \left( \frac{n_1}{\sigma_{1*}^2}, \dots, \frac{n_k}{\sigma_{k*}^2}, \frac{2n_1}{\sigma_{1*}^2}, \dots, \frac{2n_k}{\sigma_{k*}^2} \right).
\]

The test point for the PTP is determined by maximizing the profile likelihood \eqref{eqn:mean_monotone_profile} subject to the constraints $\mu_1 \leq \cdots \leq \mu_k$.
We consider a total of $2^{k-1}$ possible configurations, corresponding to whether each of the gaps $\mu_{i+1} - \mu_i$, $i\in [k-1]$, is zero or strictly positive.
Each configuration defines groups of equal means, whose values are numerically determined by maximizing \eqref{eqn:mean_monotone_profile}.
If the resulting estimates do not satisfy the monotonicity constraints $\mu_1 \leq \cdots \leq \mu_k$, the configuration is disregarded.
Among the remaining candidate parameter estimates, we select the one that yields the highest likelihood and compute the corresponding MLE of each $\sigma_i$ as
\[
  \left\{\frac{1}{n_i} \sum_{j=1}^{n_i} (x_{ij} - \mu_i)^2\right\}^{1/2}.
\]
The resulting parameter value $\theta_t = (\mu_{1,t}, \dots, \mu_{k,t}, \sigma_{1,t}, \dots, \sigma_{k,t})$ is used as the proxy test point.

The number of candidate configurations can be reduced when the sample means corresponding to the smallest indices are already monotonically ordered and are less than or equal to all remaining sample means.
For example, if $\bar x_1 \leq \bar x_2 \leq \bar x_j$ for all $j\geq 3$, then we let $\mu_{i,t} = \bar x_i$ for $i=1$ and $2$.
Similarly, we let $\mu_{k,t} = \bar x_k$ and $\mu_{k-1,t} = \bar x_{k-1}$ if $\bar x_k \geq \bar x_{k-1} \geq \bar x_j$ for all $j \leq k-2$.

For the test $H_{0,t}: \theta = \theta_t$, we use the log-likelihood ratio statistic given by
\[
  \ell(\hat \theta) - \ell(\theta_t)
  = \sum_{i=1}^k \left[ -\frac{n_i}{2} \log \frac{\hat \sigma_i^2}{\sigma_{i,t}^2} - \frac{n_i}{2} + \frac{1}{2\sigma_{i,t}^2} \left\{ n_i \hat \sigma_i^2 + n_i(\bar x_i - \mu_{i,t})^2 \right\} \right],
\]
where $\hat \theta = (\bar x_1, \dots, \bar x_k, \hat \sigma_1, \dots, \hat \sigma_k)$ denotes the unconstrained MLE, with $\hat \sigma_i = \sqrt{(n_i-1)s_i^2/n_i}$.
The null distribution of this statistic is characterized by the random variable
\begin{equation}
  \sum_{i=1}^k \left[ -\frac{n_i}{2} \log \frac{W_{i,1}}{n_i} - \frac{n_i}{2} + \frac{1}{2} W_{i,1} + \frac{1}{2} W_{i,2} \right],
  \label{eqn:mean_monotone_nulldist}
\end{equation}
where $W_{i,1} \sim \chi^2_{n_i-1}$ and $W_{i,2} \sim \chi^2_1$, $i \in [k]$, are mutually independent.

We generate $10^6$ independent draws from the distribution in \eqref{eqn:mean_monotone_nulldist} and estimate the p-value for the proxy test $H_{0,t}: \theta = \theta_t$ by the proportion of draws that are greater than the observed test statistic $\ell(\hat \theta) - \ell(\theta_t)$.
Finally, we reject $H_0: \mu_1 \leq \cdots \leq \mu_k$ if the p-value of the proxy test is less than the modified significance level $\alpha'$.

\begin{table}[t]
  \centering
  \begin{tabular}{p{3em}p{10em}p{7em}>{\raggedright}p{5em}p{5.5em}}
    \toprule
    & true means & sample sizes & true std.\linebreak deviations & rejection \linebreak probability \\ \midrule
    \multirow{3}{*}{$k=2$} & $\mathbf{(0,0)}$ & \multirow{3}{*}{$(3,5)$} & \multirow{3}{*}{$(1,2)$} & \textbf{4.9 (0.4)\%} \\
    & $\mathbf{(0,1)}$ & & & \textbf{0.6 (0.2)\%} \\
    & $(1,0)$ & & & 20.9 (0.8)\% \\ \midrule
    \multirow{7}{*}{$k=10$} & $\mathbf{(0,0,0,0,0,0,0,0,0,0)}$ & \multirow{7}{=}{$(5, 10, 15, 15, 10,\linebreak 5, 5, 10, 15, 10)$} & \multirow{7}{=}{$(3, 2, 1, 4, 2,\linebreak 1, 4, 3, 2, 1)$} & \textbf{7.4 (0.7)\%} \\
    & $\mathbf{(0,1,1,1,1,1,1,1,1,1)}$ & & & \textbf{5.7 (0.6)\%} \\
    & $\mathbf{(0,0,0,0,0,1,1,1,1,1)}$ & & & \textbf{3.8 (0.5)\%} \\
    & $(1,0,0,0,0,0,0,0,0,0)$ & & & 11.3 (0.9)\% \\
    & $(0,1,0,1,0,1,0,1,0,1)$ & & & 24.1 (1.2)\% \\
    & $\mathbf{(0,1,2,3,4,5,6,7,8,9)}$ & & & \textbf{0.0 (0.1)\%} \\
    & $(9,8,7,6,5,4,3,2,1,0)$ & & & 100.0 (0.0)\% \\ \bottomrule
  \end{tabular}
  \caption{Empirical rejection probabilities for the test for monotonicity of means.
    Cases satisfying $H_0: \mu_1 \leq \cdots \leq \mu_k$ are indicated in bold.
    Parentheses indicate 1.96 times the standard error.
}
  \label{tab:mean_monotone}
\end{table}

Table~\ref{tab:mean_monotone} shows the rejection probabilities of $H_0: \mu_1 \leq \cdots \leq \mu_k$ for $k=2$ and $k=10$ under different settings.
The true means, standard deviations, and the sample sizes for the independent normal populations are also shown.
The target significance level is $\alpha = 0.05$.
The empirical rejection rates are close to or below $\alpha$ when the true mean vector $\mu = (\mu_1, \dots, \mu_k)$ satisfies $H_0$, whereas they are greater than $\alpha$ when $\mu$ does not satisfy $H_0$.
When $\mu_1 = \cdots = \mu_k$, that is, when the true parameter lies on the boundary of $\Theta_0$, the empirical rejection rates are reasonably close to the target significance level, even for relatively small sample sizes.
The difference between the empirical rejection rates and the significance level decreases as the standard deviations decrease or the sample sizes increase (results not shown).

\section{Discussion}\label{sec:discussion}

We developed a method for testing composite null hypotheses at a specified significance level.
The null hypothesis is rejected, in principle, if every point in the null region is rejected at a modified significance level $\alpha'$.
In practice, the rejection decision can be based on a finite number of test points, and our procedure often allows a single test point to be used.
We developed a general formula for determining $\alpha'$ for composite null hypotheses defined by an arbitrary number of equality and inequality constraints.
Our approach can maintain a Type I error rate close to the nominal significance level in small-sample settings by using finite-sample methods for proxy hypothesis tests.
The PTP substantially generalizes the likelihood ratio test by allowing null regions defined by inequality constraints as well as equality constraints.

Through several examples, we demonstrated that our method provides accurate tests for nonstandard composite hypotheses.
It also provides a principled finite-sample framework for hypothesis testing and confidence interval construction in the presence of nuisance parameters.

Although this paper focuses on parametric inference, the idea of the PTP may also be applied to nonparametric tests of functionals of the underlying probability distribution.
For a functional $\lambda(F)$, such as the mean or a quantile, the composite null $H_0: \lambda(F) \in A_0$ may be tested by testing the simple null hypothesis $H_0: \lambda(F) = \lambda_0$ for each $\lambda_0 \in A_0$ at a modified significance level.
These simple hypotheses may be tested using nonparametric confidence intervals, including bootstrap-based or empirical likelihood ratio confidence intervals \citep{efron1981, efron1987, owen1988}.
The development of nonparametric PTP is left for future work.

\section{Acknowledgments}
\if1\anon
Joonha Park gratefully acknowledges support from the General Research Fund of the College of Liberal Arts and Sciences at the University of Kansas and from the Don and Pat Morrison Foundation.
\fi

\section{Disclosure statement}\label{disclosure-statement}
The authors report that there are no competing interests to declare.
The authors used the following generative AI tools in the preparation of this manuscript: ChatGPT (versions GPT-4o and GPT-5) for language improvement, and Gemini (versions gemini-2.5-flash and gemini-2.0-flash-001) for coding assistance. All content generated or assisted by these tools was carefully reviewed and revised by the authors to ensure accuracy and appropriateness.

\section{Data Availability Statement}\label{data-availability-statement}

The data that support the findings of this study are openly available in 4TU.ResearchData at http://doi.org/10.4121/03d6b167-2e33-43dc-ada7-40c225094ada, with reference number Ouqh6LTmNowf-hIpFEZ5XoWWS8eoLWpC3t6Q0Q55Z5A.

\bibliography{references}

\includepdf[pages=-]{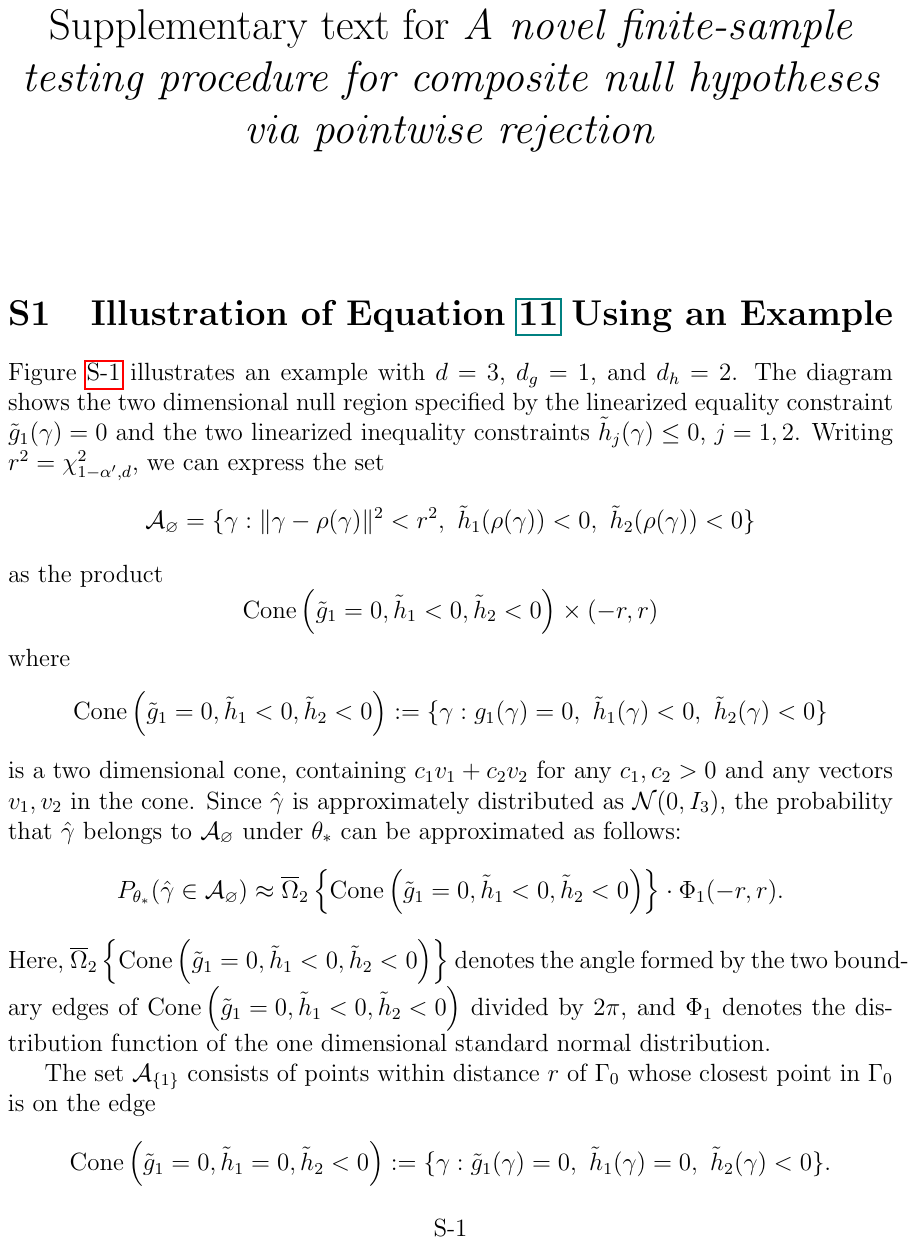}

\end{document}